\documentclass[preprint,1p]{elsarticle}
\usepackage{multicol}
\usepackage{amssymb,amsmath}
\usepackage{longtable}
\usepackage{dcolumn} % Align table columns on decimal point
\usepackage{bm} % bold math
\usepackage{float}
\usepackage{setspace,supertabular}
\usepackage{lineno}
 \usepackage{subfigure}
\usepackage{graphicx}

\usepackage[usenames,dvipsnames]{color}
\biboptions{sort&compress}

\DeclareFontFamily{OT1}{pzc}{}
\DeclareFontShape{OT1}{pzc}{m}{it}%
              {<-> s * [0.900] pzcmi7t}{}
\DeclareMathAlphabet{\mathpzc}{OT1}{pzc}%
                                 {m}{it}

\newcommand{\vect}[1]{\boldsymbol{#1}}

\begin{document}

\begin{frontmatter}

\title{Relativistic Three-Dimensional Lippmann-Schwinger Cross Sections for Space Radiation Applications}% Force line breaks with \\

\author[LaRC]{C.~M.~Werneth}\corref{cor1}
\ead{charles.m.werneth@nasa.gov}
\author[NIA]{X.~Xu}%
\author[LaRC]{R.~B.~Norman}
\author[USM]{K.~M.~Maung}

\cortext[cor1]{Corresponding Author}
\address[LaRC]{%
 NASA Langley Research Center, United States
}
\address[NIA]{%
National Institute of Aerospace, United States
}

\address[USM]{%
The University of Southern Mississippi, United States
}

\date{\today}

\begin{abstract}
Radiation transport codes require accurate nuclear cross sections to compute particle fluences inside shielding materials. The Tripathi semi-empirical reaction cross section, which includes over 60 parameters tuned to nucleon-nucleus (NA) and nucleus-nucleus (AA) data, has been used in many of the world's best-known transport codes. Although this parameterization fits well to reaction cross section data, the predictive capability of any parameterization is questionable when it is used beyond the range of the data to which it was tuned. Using uncertainty analysis, it is shown that a relativistic Three-Dimensional Lippmann-Schwinger (LS3D) equation model based on Multiple Scattering Theory (MST) that uses 5 parameterizations---3 fundamental parameterizations to nucleon-nucleon (NN) data and 2 nuclear charge density parameterizations---predicts NA and AA reaction cross sections as well as the Tripathi cross section parameterization for reactions in which the kinetic energy of the projectile in the laboratory frame ($T_{\rm Lab}$) is greater than 220 MeV/n.  The relativistic LS3D  model has the additional advantage of being able to predict highly accurate total and elastic cross sections. Consequently, it is recommended that the relativistic LS3D model be used for space radiation applications in which $T_{\rm Lab}>$ 220 MeV/n.  
\end{abstract}

\end{frontmatter}

%\maketitle
%%\linenumbers

	Radiation transport codes describe the propagation of particles and secondary particles produced from their collisions from a specified boundary to a location inside a shielding material or body to obtain particle flux, which can then be used to estimate dosimetric quantities.  Some of the transport codes employ semi-empirical parameterizations to the available cross section data to describe physical interactions. The \citet{Tripathi} reaction cross section model for nucleon-nucleus (NA) and nucleus-nucleus (AA) reactions is implemented in many of the world's state-of-the-art radiation transport codes \cite{Geant4,Phits1,Phits2,Slaba1,Slaba2}. This parameterization fits well to the available data \cite{Werneth_Validation}, but the predictive capability of any parameterization is suspect beyond the range of the data to which it was tuned. Therefore, physical models are preferred, provided that the fidelity is acceptable as compared to available data, since theoretical predictions are expected to be more accurate than parameterizations when used for reactions beyond the range of the fitted data.

	In a recent study focused on space radiation applications (projectile kinetic energies in the lab frame ($T_{\rm Lab}$) up to 100 GeV/n for nuclei with masses extending up to Nickel), \citet{Werneth_Validation} used uncertainty quantification (UQ) to compare elastic differential, elastic, reaction, and total cross section predictions of three models: the Eikonal (Eik) approximation, the partial wave (PW) solution method of the Lippmann-Schwinger equation, and the three-dimensional Lippmann-Schwinger (LS3D) equation.  The PW and LS3D solution methods were expressed in the momentum-space representation to allow for the incorporation of relativistic kinematics, and as expected, the relativistic models were in better agreement with experimental data.  The Tripathi parameterization \cite{Tripathi}, however, was shown to be in better agreement with the reaction cross section data than the Eik, PW, and LS3D models. Based on this study, it was recommended that the relativistic LS3D model should be used for total, elastic, and elastic differential cross sections for space radiation applications when the projectile kinetic energy in the laboratory frame was greater than 220 MeV/n, whereas the Tripathi parameterization should be used for all reaction cross sections.  Based on the present analysis it is proposed that this recommendation be changed.
	
The Eik, PW, and LS3D models are based on the Multiple Scattering Theory (MST) of interaction in which the physical observables are computed from the scattering transition matrix \cite{WernethPRC,Werneth_Eqmass},
	\begin{equation}
		T(\vect k',\vect k) = U(\vect k',\vect k) + \int U(\vect k', \vect k'') G_0(k,k'') T(\vect k'', \vect k) d\vect k'',
	\end{equation}
where $\vect k$ ($\vect k'$) is the incident (final) relative momentum of the projectile-target system in the center of momentum frame, $U(\vect k', \vect k'')$ is the optical potential, and $G_0(k,k'')$ is the two-body free propagator. Using the impulse, optimum factorization, and on-shell approximations, the optical potential may be expressed as \cite{Werneth_Validation,Wolfe}
\begin{equation}
	 U(\vect k', \vect k) \equiv U(|\vect q|) = A_P A_T t_{\rm NN}(q)\rho_P(q) \rho_T(q),
	\end{equation}
where $A$ is the number of nucleons in the projectile ($P$) and target ($T$), $\rho(q)$ is the nuclear charge density, $t_{\rm NN}(q)$ is the NN transition amplitude, and $q =|\vect q|$ is the momentum transfer. The transition amplitude is modeled as a Gaussian function of momentum-transfer and consists to three energy dependent parameters: NN slope parameter, total NN cross section, and the real-to-imaginary ratio of the NN transition amplitude, which have been provided explicitly in previous work \cite{Werneth_Validation}.

Harmonic-Well (HW) nuclear charge densities were used for lighter nuclei ($A\le16)$, where the charge density profile may be approximated with a Gaussian-like distance fall-off, and heavier nuclei charge densities were modeled with the two-parameter Fermi (2PF) profiles. Charge density parameters were extracted from electron-nucleus scattering experiments \cite{DeVries1,DeVries2}.  When HW charge densities were not provided, an isotopic parameter average was performed \cite{Werneth_Validation}. The 2PF charge density parameters were estimated with the Nuclear Droplet model \cite{Myers}.  

At relativistic energies, it is expected that the internal quark structure of the nucleons will be probed. Since the fundamental constituent particles must be specified in any MST, it is assumed  that that individual nucleons are not composite particles and that any relativistic effects associated with probing the nucleons at these energies have been absorbed into the parameters. In order to account for this assumption, the nuclear charge densities must be converted to matter densities. 
This was accomplished for both the HW and 2PF models by using the approach described elsewhere 
\cite{Werneth_Validation,WilsonTownsend,NASATN,NASARP1257}.  In the momentum space representation the matter density may be obtained by using \cite{NASARP1257}
\begin{equation}
   \rho_{m}(q) = \rho_c(q)/\rho_p(q) \label{rhoq},
\end{equation}
where $\rho_m(q)$ is the matter density, $\rho_c(q)$ is the charge density, and $\rho_p(q)$ is the proton charge density. Although in previous studies, the calculations of the nuclear cross sections were performed in the position-space representation \cite{WilsonTownsend,NASATN,NASARP1257}, the HW charge density was transformed to momentum-space in current analysis, and equation (\ref{rhoq}) was employed to find the matter density . Next, the HW nuclear matter density was transformed back to position-space, where the calculation was performed. 

In references \cite{WilsonTownsend,NASATN}, the 2PF matter density is unfolded in the position space representation with a 2 point Gaussian-Hermite quadrature formula, which yields nuclear matter densities that have the same 2PF profile as the nuclear charge density parameterization but with modified parameters. This approach is convenient, but it will be shown that a different proton charge density correction evaluated completely in the momentum-space representation yields significantly better results.

The current work implements a proton charge density of the following form for heavy ions:
\begin{equation}
\rho_p(q) = \frac{1}{[1 + (q^2 a_0^2)]^2}, \label{newparam}
\end{equation}
which provides an equally good fit to the data as the Gaussian parameterization \cite{Hofstadter1958}. There are numerical advantages to using this particular parameterization as compared to a Gaussian fit. When utilized for the heavy ion nuclear matter density, the Gaussian fit of the proton charge density has a tendency to decay rapidly to small values, and approximations are needed to ensure the proper behavior of equation (\ref{rhoq}) at large momentum transfer. The parameterization in equation (\ref{newparam}) does not exhibit this behavior and leads to proper convergence without further approximations. The parameter reported by Wolfe et al. \cite{Wolfe,Pickle} was implemented in the present work: $a_0 = 1.0/18.77$ fm$^2$.

The uncertainty quantification (UQ) metrics described in \cite{NormanJCP,NormanTP} were used to compare the previous results by \citet{Werneth_Validation} to the new results presented herein. The UQ analysis treats each experimental datum point as an interval defined by its uncertainty. Since the experimental measurement is represented by an interval, two measurements are defined: (1) the minimum difference between the model and uncertainty interval and (2) the maximum difference between the model and the uncertainty interval. The maximum absolute difference represents the worst case, absolute uncertainty (AU) metric and is useful for applications in which the magnitudes of the differences are important. Another UQ metric employed in this analysis is the relative uncertainty (RUC), which is a measure of the relative difference between the model and experiment and is useful for validation of model predictions.  Probability distribution functions (PDFs) are assembled for both UQ metrics, and the median of the Cumulative Distribution Function (CDF) is reported herein. See references \cite{Werneth_Validation,NormanJCP,NormanTP} for more discussion and justification of the UQ metrics.  

 In the following figures, the theoretical predictions of the Eik, non-relativistic (NR) and relativistic (REL) LS3D models, and the Tripathi parameterizations are compared to experimental data. Eik results are shown in green, LS3D (NR) in blue, LS3D (REL) in red, Tripathi in pink, and experimental data in black. Smaller medians are indicative of better agreement of models with experimental data.  Figs. 1 and 2 include all total, elastic, and reaction cross section measurements that are available in the database \cite{Werneth_Validation}, and Figs. 3 and 4 include only reaction cross section comparisons. The number of data points in each each energy bin is given by ``npt''. The results of reference \cite{Werneth_Validation} are shown with hashed bars and were generated with the Gauss-Hermite approximation.  The current results are shown with solid bars and were produced with the proton charge density given in equation (\ref{newparam}).  The energy bins were changed slightly from those previously reported \cite{Werneth_Validation}. Reactions with $T_{\rm Lab} <$ 220 MeV/n were excluded since the models are not well-suited for such low energies without the inclusion of medium effects in the MST. In the previous study, very little experimental data were available in the 5-10 GeV/n energy bin, so these measurements have been included in the ``$>$5 GeV/n'' energy bin. 
	
Figs. 1-4 show a significant improvement in the agreement of the theoretical models with experiment as compared to previous efforts \cite{Werneth_Validation}. The AU metric in Fig. 1 shows that predictions are now roughly 50 mb closer to experimental measurements, whereas Fig. 2. shows that the RUC\% has decreased in every energy bin. Figs. 3 and 4 show that the LS3D (REL) model can predict reaction cross sections equally well as the Tripathi parameterization \cite{Tripathi}, which uses over 60 parameters that are tuned to NA and AA data. The differences between the LS3D (REL) and Tripathi results are smaller than the experimental uncertainty.  

The UQ analysis shows that the REL LS3D model predicts reaction cross sections as well as the Tripathi parameterization, and it also predicts highly accurate total and elastic cross sections. Consequently, it is recommended that the LS3D (REL) model be used for space radiation applications for $T_{\rm Lab} > $ 220 MeV/n. 

This work was supported by the Human Research Program under the Human Exploration and Operations Mission Directorate
of NASA and NASA Grant No. NNX13AH31A. This work was also by supported by contributions from the National Institute of Aeorspace.

%\bibliographystyle{apsrev}
%\bibliography{scat_Wolfe}
\clearpage

%%%%%%%%%%%%%%%%%%%%%%% FIGURES %%%%%%%%%%%%%%%%%%%%%%%%%%%%%%%%%%%%%%%%%%%%

%%%% FOUR ENERGY GROUPS

\begin{figure}
	\centering
		\includegraphics[scale = 0.6]{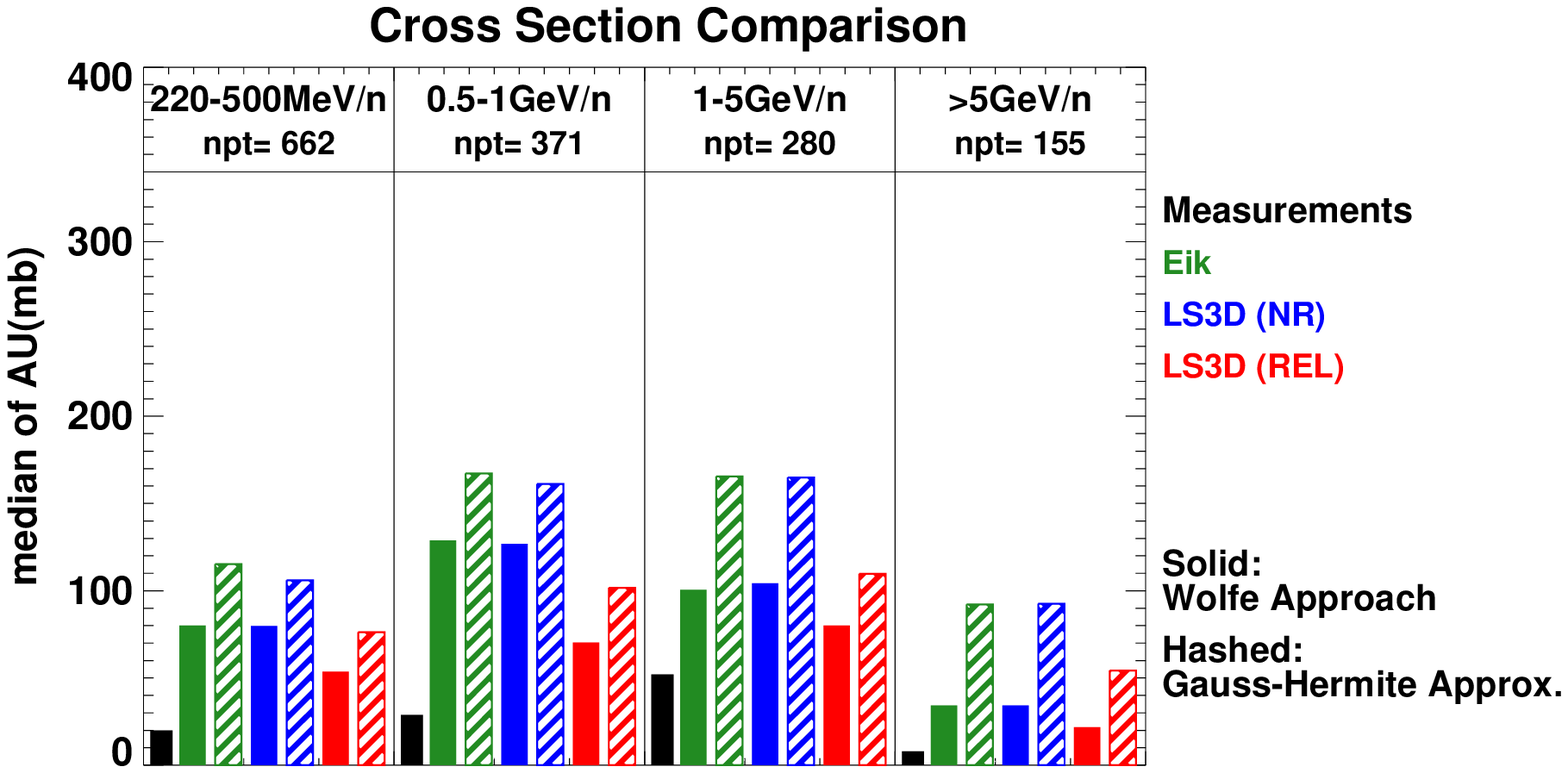}
	\caption[fig1]{Comparisons of total, elastic, and reaction cross sections utilizing the Eik, PW, and LS3D models with the median of the absolute uncertainty cumulative distribution function. Smaller medians indicate better agreement with experimental data \cite{Werneth_Validation}.}
	\label{fig:AUyayc4}
\end{figure}

\begin{figure}
	\centering
		\includegraphics[scale = 0.6]{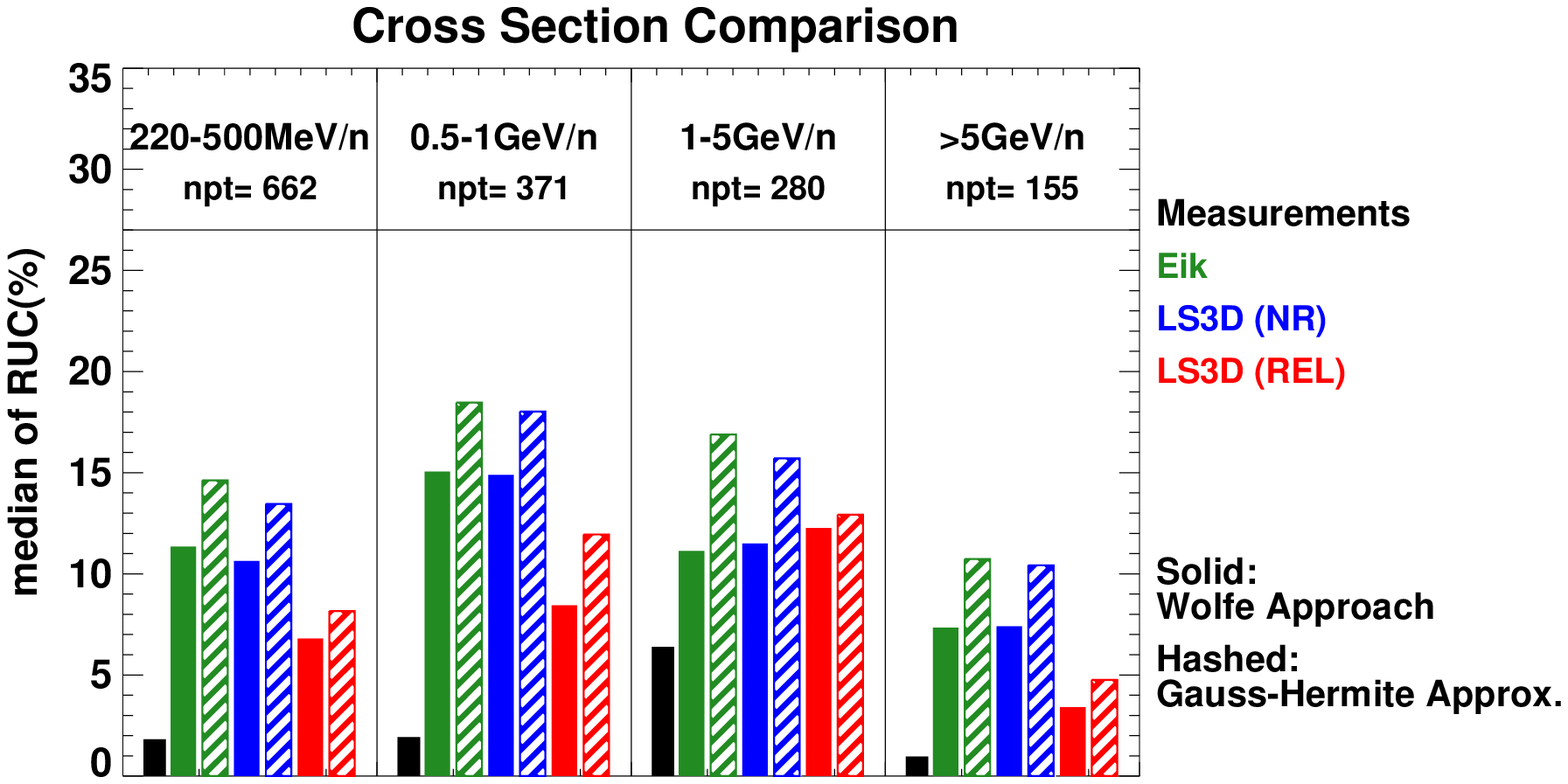}
	\caption[fig2]{Comparisons of total, elastic, and reaction cross sections utilizing the Eik, PW, and LS3D models with the median of the relative uncertainty distribution function. Smaller medians indicate better agreement with experimental data \cite{Werneth_Validation} }
	\label{fig:RUCyayc4}
\end{figure}

\clearpage
% Reaction cross section bar graphs
\begin{figure}
	\centering
		\includegraphics[scale = 0.6]{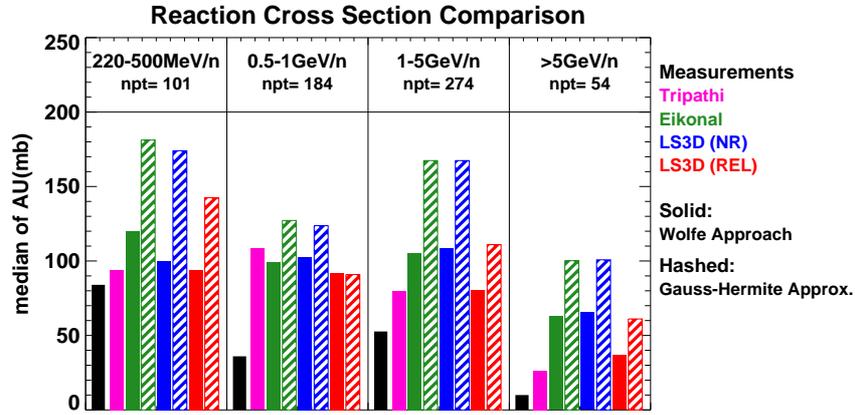}
	\caption[fig3]{Comparisons of reaction cross sections utilizing the Eik, PW, LS3D, and Tripathi models. The median of the AU cumulative distribution function has been used as the UQ metric. Smaller medians indicate better agreement with experimental data \cite{Werneth_Validation} .}
	\label{fig:reactionAU4}
\end{figure}

\begin{figure}
	\centering
		\includegraphics[scale = 0.6]{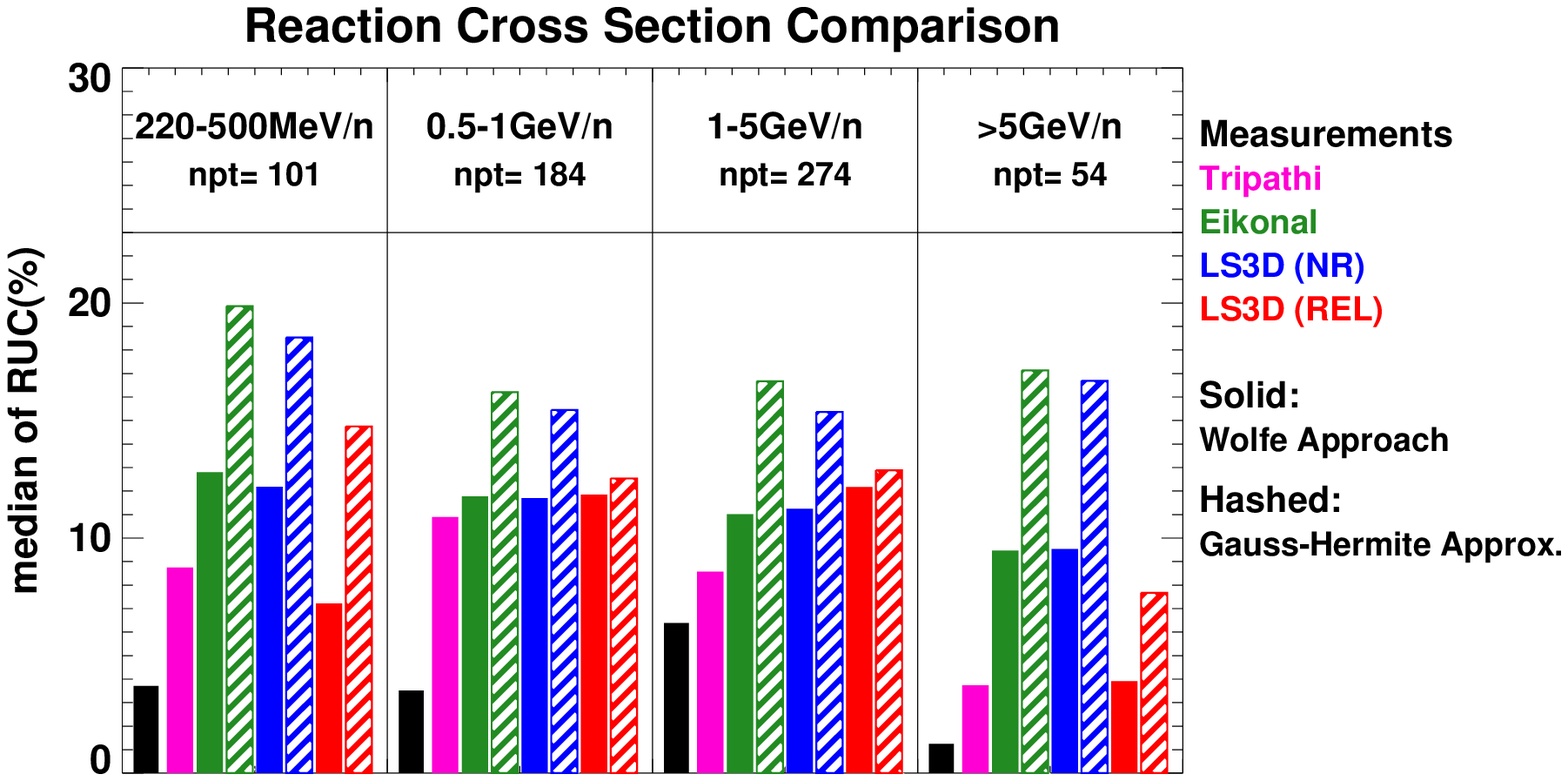}
	\caption[fig4]{Comparisons reaction cross sections utilizing the Eik, PW, and LS3D models with the median of the relative uncertainty distribution function. Smaller medians indicate better agreement with experimental data \cite{Werneth_Validation}.}
	\label{fig:reactionRUC4}
\end{figure}

\end{document}